\documentclass[a4paper,12pt]{article}
\usepackage{amssymb}

\def\lbeq#1{\begin{equation} \label{#1}} 
\def\eeq{\end{equation}} 
\def\bary{\begin{array}}
\def\eary{\end{array}}
\def\gzit#1{{\rm (\ref{#1})}} 			
\def\fct#1{\mathop{\rm #1}}	
\def\fns#1{{\mbox{\rm \scriptsize#1}}} 		
\def\D{\displaystyle}				
\def\newl{\hfill\break}				
\def\wave{\protect{\footnotesize $\sim$}}
\def\pdot{{\raisebox{-.2em}{\hbox{\LARGE $\displaystyle \cdot$}}}}
\def\<{\langle} 				
\def\>{\rangle} 				
\def\half{\frac{1}{2}} 
\def\downto{\downarrow}
\def\hbar{{\mathchar'26\mkern-9muh}}
\def\lp{\mbox{\Large$\,_\urcorner\,$}}   
\def\slp{\,_\urcorner\,}                 
\def\sint{\mbox{$\int$}}
\def\forall{~~~\mbox{for all }}
\def\cons{\fns{cons}}
\def\Diag{\fct{Diag}}
\def\Lin{\fct{Lin}~}
\def\poly{\fns{pol}}
\def\re{\fct{Re}}		

\def\tr{\fct{tr}}
\def\eps{\varepsilon}
\def\0 {{\bf 0}}
\def\k {{\bf k}}
\def\p {{\bf p}}
\def\q {{\bf q}}
\def\r {{\bf r}}
\def\uu {{\bf u}}
\def\x {{\bf x}}
\def\Cz{\mathbb{C}}
\def\Ez{\mathbb{E}}
\def\Hz{\mathbb{H}}
\def\Lz{\mathbb{L}}
\def\Rz{\mathbb{R}}
\def\Dc{{\cal D}}

\advance\textwidth2.0cm         
\advance\evensidemargin-0.9cm
\advance\oddsidemargin-0.9cm

\makeindex

\begin{document}


\begin{center}

{\LARGE\bf Quantum field theory as eigenvalue problem} \\

\vspace{1cm}

\centerline{\sl {\large \bf Arnold Neumaier}}

 \vspace{0.2cm}

\centerline{\sl Institut f\"ur Mathematik, Universit\"at Wien}
\centerline{\sl Strudlhofgasse 4, A-1090 Wien, Austria}
\centerline{\sl email: Arnold.Neumaier@univie.ac.at}
\centerline{\sl WWW: http://www.mat.univie.ac.at/\wave neum/}
\end{center}

\hfill March 10, 2003

{\small
{\bf Abstract.} 
A mathematically well-defined, manifestly covariant theory of 
classical and quantum field is given, based on Euclidean Poisson 
algebras and a generalization of the Ehrenfest equation, which
implies the stationary action principle. 
The theory opens a constructive spectral approach to finding 
physical states both in relativistic quantum field theories and for
flexible phenomenological few-particle approximations. 

In particular, we obtain a Lorentz-covariant phenomenological 
multiparticle quantum dynamics for electromagnetic and gravitational 
interaction which provides a representation of the 
Poincar\'e group without negative energy states. The dynamics 
reduces in the nonrelativistic limit to the traditional Hamiltonian 
multiparticle description with standard Newton and Coulomb forces.

The key that allows us to overcome the traditional problems in 
canonical quantization is the fact that we use the algebra of 
linear operators on a space of wave functions slightly bigger 
than traditional Fock spaces.

\bigskip
\begin{flushleft}
{\bf Keywords}: 
action,
axiomatic physics,
classical field,
classical limit,
conservative quantum state,
constrained Schr\"odinger equation,
covariant interaction,
deformation quantization,
density,
multiparticle Dirac equation,
Ehrenfest equation,
eigenvalue problem,
electrodynamics,
Euclidean Poisson algebra,
expectation,
form factors,
gravitation,
Hamiltonian dynamics,
integral,
Liouville equation,
manifestly covariant theory,
momentum state,
noncommutative algebra,
nonrelativistic limit,
phase space quantization,
phenomenological relativistic dynamics,
physical system,
quantum field theory,
quantum gravity,
relativistic Coulomb potential,
relativistic Newton potential,
running coupling constants,
self-energy,
stationary action principle,
Wightman axioms,
Wigner transform
\end{flushleft}

{\bf E-print Archive No.}: gr-qc/0303037
   \newl
{\bf 2003 PACS Classification}: 03.70, 04.60.-m, 11.10.Cd, 11.10.Ef
   \newl
}


\newpage
\section{Introduction} \label{intro}

\hfill\parbox[t]{11.8cm}{\footnotesize 

{\em 
\ldots the ancients (as we are told by Pappus) esteemed the science of 
mechanics of greatest importance in the investigation of natural 
things, and the moderns, rejecting substantial forms and occult 
qualities, have endeavored to subject the phenomena of nature to the 
laws of mathematics \ldots
}

Isaac Newton, 1686 \cite{New}

\bigskip
{\em
Renormalized quantum electrodynamics is by far the most successful 
theory we have today. This very impressive fact, however, does not 
make the whole situation less strange.
We start out from equations which do not make sense. We apply certain 
prescriptions to their solutions and end up with a power series of 
which we do not know that it makes sense. The first few terms of 
this series, however, give the best predictions we know.
}

Res Jost, 1965 \cite{Jos}
}\nopagebreak

In the more than 300 years that passed since Newton wrote this in his
{\em Principia Mathematica}, the moderns have been very successful 
at the endeavor to subject the phenomena of nature to the laws of 
mathematics -- with exception of quantum field theory. As the second 
quote (which could have as well been written in 2002) shows, 
quantum field theory so far resisted a quantitative, mathematically 
rigorous foundation. 

In the present paper, an axiomatic approach is outlined that, I believe,
provides foundations on which quantum field theory can be given a 
rigorous mathematical treatment. The present paper gives the elementary 
part and exhibits the connections to the traditional settings. 
A deeper study of the consequences and use of the concepts presented 
here will be given elsewhere. 

In the new approach, each (classical or quantum) conservative 
physical system is characterized by two Hermitian quantities: 
a density and an action.
A generalized Liouville equation defines the dynamics and implies
Ehrenfest equations for expectations.

For a classical (but not a quantum) field theory, the 
Ehrenfest equations in a symplectic Poisson algebra imply the 
traditional field equations by the stationary action principle. 
In particular, all traditional systems derivable from the stationary 
action principle can be modelled in our setting. In a similar way,
one can get from suitable Lie-Poisson algebras relativistic and 
nonrelativistic Euler equations, Vlasov-Maxwell, Vlasov-Einstein, 
and Euler-Poincar\'e equations.

A new phase space quantization principle (generalizing the Wigner
transform) allows the simple quantization of arbitrary Poisson 
algebras, with a good classical limit.

A large class of Poincar\'e invariant actions on spaces with a 
reducible representation of the Poincar\'e group is exhibited.
Since it is manifestly covariant but possesses a Hamiltonian 
nonrelativistic limit, it appears to be well-suited for 
phenomenological modeling of relativistic few-particle dynamics. 

In particular, we obtain a Lorentz-covariant phenomenological 
multiparticle quantum dynamics for electromagnetic and gravitational 
interaction which reduces in the nonrelativistic limit to the 
traditional Hamiltonian multiparticle description with standard 
Newton and Coulomb forces.
The key that allows us to overcome the traditional problems in 
canonical quantization is the fact that we use the algebra of 
linear operators on a space of wave functions slightly bigger 
than traditional Fock spaces.

For a quantum system, if the action is translation invariant,
one can find pure states of given mass describing isolated systems 
in a rest frame by solving a constrained Schr\"odinger equation. 
This opens a constructive spectral approach to finding 
physical states both in relativistic quantum field theories and in
phenomenological few-particle approximation.

While I have already checked much of what is
needed to get the many known results as consequences of the present
setting, I am not yet completely sure about the adequacy of the new 
theory for all aspects of traditional field theory. Thus I'd like to 
apologize (as Newton did in the preface of \cite{New}) and 
{\em ``heartily beg that what I have here done may be read with 
forbearance; and that my labors in a subject so difficult may be 
examined, not so much with the view to censure, as to remedy their 
defects.''}

\section{Prelude: Covariant transmutation}

\hfill\parbox[t]{11.3cm}{\footnotesize 

{\em 
Do not imagine, any more than I can bring myself to imagine, 
that I should be right in undertaking so great and difficult a task.
Remembering what I said at first about probability, I will do my best 
to give as probable an explanation as any other -- or rather, more 
probable; and I will first go back to the beginning and try to speak 
of each thing and of all.
}

Plato, ca. 367 B.C. \cite{Pla}

}\nopagebreak

We begin with traditional nonrelativistic quantum mechanics of a
multiparticle system and
rewrite it in a formally covariant way that foreshadows the axiomatic 
setup developed afterwards. (This section serves as a heuristic 
motivation only, without any claims to rigor.) 

Let $H$ be a translation invariant and time-independent Hamiltonian,
$\p$ the 3-momentum operator, the generator of the spatial translations,
$\psi$ the energy eigenstate in the 
rest frame of a system with energy $E>0$ and total mass $m$.
The Schr\"odinger equation gives $H\psi=E\psi$, and the condition that
the system is in a rest frame says $\p\psi=0$.

To make these statements covariant, we extend wave functions by
an additional argument $E$. Then we may consider the energy $E$ 
as an operator acting on functions $\psi(E,x)$ by multiplication 
with $E$, with time $t=i\hbar\partial/\partial E$ as conjugate
operator. By introducing the operator 
\lbeq{e.action}
L:=E-H,
\eeq
we may write the Schr\"odinger equation together with the rest frame 
condition in the form
\[
L\psi=0,~~~\p\psi=0.
\]
We now introduce a 4-momentum vector $p={p_0 \choose \p}$, where
$p_0$ is related to the energy $E$ by the relation
\lbeq{e.nrenergy}
E=p_0c-mc^2.
\eeq
On writing $p^2=p_0^2-\p^2$ (where $\p^2=\p\cdot\p$)
for the Lorentz square of a 4-vector, 
and applying a Lorentz transform, we see that this is 
equivalent with the condition 
\[
L\psi=0,~~~
p\psi=k\psi
\]
for a pure momentum state $\psi$ with definite
4-momentum $k>0$ (in the forward cone, i.e., $k_0>|\k|$) and energy 
\[
E=c\sqrt{k^2}-mc^2.
\]
A general pure state is a superposition $\psi=\int dk \psi_k$ of
(unnormalized) momentum states with 4-momentum $k$, 
hence any nonzero $\psi$ with 
\[
L\psi=0,~~~\mbox{i.e.,}~~~\psi\in \Hz:=\ker L.
\]
A general mixed state is a mixture
$\rho=\D\int d\pi(\alpha)\psi_\alpha\psi_\alpha^*$ of pure states 
$\psi_\alpha$ 
weighted by a nonnegative measure $d\pi(\alpha)$. 
Thus $L\rho=\rho L=0$. In particular, with the standard expectation
\[
\<f\>_\rho:=\tr \rho f,
\]
of linear operators $f$ on functions $\psi(E,x)$
we have
\lbeq{e.start}
[L,\rho]=0,~~~\<L\>_\rho=0.
\eeq
The equations \gzit{e.start} will be the starting point for our 
axiomatic setting. Slightly generalized, they will allow us to 
formulate not only nonrelativistic quantum mechanics, but also 
classical mechanics, classical field theory, and quantum field theory.

\section{Axiomatic physics}

\hfill\parbox[t]{12.1cm}{\footnotesize 

{\em
Das Streben nach Strenge zwingt uns eben zur Auffindung einfacherer 
Schlu{\ss}\-weisen; auch bahnt es uns h\"aufig den Weg zu Methoden, die 
ent\-wickelungsf\"ahiger sind als die alten Methoden von geringerer 
Strenge. [...]
\\
Durch die Untersuchungen \"uber die Grundlagen der Geometrie
wird uns die Aufgabe nahe gelegt, nach diesem Vorbilde diejenigen 
Disziplinen axiomatisch zu behandeln, in denen schon heute die 
Mathematik eine hervorragende Rolle spielt; dies sind in erster 
Linie die Wahrscheinlichkeits\-rechnung und die 
Mechanik.
}

David Hilbert, 1900 \cite{Hil}

}\nopagebreak

We now begin the axiomatic treatment; from now on, all concepts have 
a precise, unambiguous meaning. Here we concentrate on the conservative,
(classical and quantum) mechanics part of Hilbert's 6th problem, 
quoted above; for the probability part, viewed in the present context, 
see {\sc Neumaier} \cite{Neu.ens}. In this paper, we only give the 
outlines and general flavor of the theory. A much more extensive 
version with full details, and a treatment of the dissipative case 
are in preparation. 

The {\bf quantities} of interest are elements of a 
{\bf Euclidean Poisson algebra} $\Ez$ containing the complex numbers
as {\bf constants}. 
Apart from an associative product (commutative only in the 
classical case) one has an {\bf involution} $*$ reducing on the 
constants to complex conjugation, a complex-valued {\bf integral} 
$\int$ defined on a subalgebra $I\Ez$ of {\bf integrable} quantities, 
and a {\bf Lie product} (or {\bf bracket}) $\lp$. The subalgebra 
$B\Ez$ of {\bf bounded} quantities consists of all $f\in\Ez$ with 
$f^*f\le \alpha^2$ for some $\alpha\in\Rz$.
Quantities $f$ with $f^*=f$ are called {\bf Hermitian}.
(For reasons given in {\sc Neumaier} \cite{Neu.ens}, we avoid using 
the customary word `observables', and follow instead the 
International System of Units (SI) \cite{SI} in our terminology.)

Apart from the standard rules for $*$-algebras and the linearity of 
the integral and the Lie product, one assumes the following axioms. 
(The product has priority 
over the Lie product, and both have priority over the integral.
The partial order is defined by $f\ge 0$ iff $f^*=f$ and 
$\sint g^*fg\ge0$ for all $g\in I\Ez$, 
and the monotonic limit is defined by
$f_l \downarrow 0$ iff, for every $g \in I\Ez$, the sequence 
(or net) $\sint g^*f_lg$ consists of real numbers converging 
monotonically to zero.) 

{\bf Axioms for a Euclidean $*$-algebra:}\\
(E1)~ $f \in B\Ez,~ g\in I\Ez ~~\Rightarrow~~ g^*,fg,gf \in I\Ez$\\
(E2) ~$(\sint g) ^* = \sint g^*, ~~~\sint fg = \sint gf$\\
(E3) ~$ \sint g^* g > 0$ ~if $g \not= 0$\\
(E4) ~$\sint g^* f g= 0$ for all  $g \in I\Ez ~~\Rightarrow~~ f=0$~~~
{\bf (nondegeneracy)}\\
(E5) ~$ \sint g_l^* g_l \to 0 ~~\Rightarrow~~ \sint f g_l \to 0$,~
$\sint g_l^* f g_l \to 0$\\
(E6) ~$g_l\downto 0~~\Rightarrow~~ \inf\sint g_l=0$~~~
{\bf (Dini property)}\\

{\bf Additional axioms for a Euclidean Poisson algebra:}\\
(P1) ~$(f \lp g)^* =f^* \lp g^*$\\
(P2) ~$f \lp g = - g \lp f $ ~~~ 
{\bf (anticommutativity)}\\
(P3) ~$f \lp(g\lp h)=(f\lp g)\lp h+g\lp(f\lp h)$~~~
{\bf (Jacobi identity)}\\
(P4) ~$ f \lp gh = (f \lp g) h + g(f \lp h)$~~~
{\bf (Leibniz identity)}\\
(P5) ~$f^* f = 0 ~~\Rightarrow~~ f = 0$~~~
{\bf (nondegeneracy)}\\
(P6)~ $f\in I\Ez,~g\in \Ez ~~\Rightarrow~~ f \lp g \in I\Ez$,\\
(P7)~ $\sint f \lp g =0$ if $f\in I\Ez$~~~
{\bf (partial integration)}\\
As a consequence,\\
(P8)~ $\sint f(g\lp h)=\sint (f\lp g)h$.

Note that (E3) implies the Cauchy-Schwarz inequality
\[
\sint (fg)^*(fg)\le \sint f^*f~\sint g^*g,
\]
which implies thast $I\Ez$ is contained in $B\Ez$.

The present definition of a Euclidean Poisson algebra is a 
modification of the concept of a Poisson algebra
as discussed in {\sc Vaisman} \cite{Vai} and 
{\sc da Silva \& Weinstein} \cite{daSW} in that commutativity is 
dropped but integration requirements imposing a Euclidean structure
are added. This modification enables us to treat classical and 
quantum physics on the same footing. (For a related attempt in this
direction, see {\sc Landsman} \cite{Lan}.)
Moreover, we introduced the symbol $\lp$ (an inverted 
stylized L, read 'Lie') to replace the Poisson bracket notation, 
which would be much more cumbersome if used extensively (as in as 
yet unpublished work).

In the present, elementary paper, we make use only of some of the 
above axioms (mainly those not involving limits). However, as will 
be shown elsewhere, all are needed for the deeper analysis of our 
conceptual basis.

To show that the axioms are rich in contents, we describe two basic 
realizations of them.

\bigskip
{\bf The quantum Poisson algebra.}
Let $\Hz$ be a Euclidean (= pre-Hilbert) space.
We define the commutator $[f,g]:=fg-gf$,
and let $\iota:=i/\hbar$ with a positive real number $\hbar$ called 
{\bf Planck's constant}.
Then the algebra $\Ez = \Lin \Hz$ of continuous linear operators 
on $\Hz$ is a Poisson algebra with {\bf quantum bracket}  
\[
  f \lp g = \iota [f,g],
\]
and Euclidean with {\bf quantum integral}
\[
  \sint f= \tr f,
\]
Integrable quantities are the operators $f\in\Ez$ for which all $gfh$
with $g,h\in\Ez$ are trace class. (This includes all operators of 
finite rank.) The axioms are easily verified. 

\bigskip
{\bf Nonrelativistic quantum mechanics.}
Nonrelativistic quantum physics is usually described by a rigged 
Hilbert space
(see, e.g., {\sc Bohm} \cite{Boh}),
if one wants to have direct access to the unbounded operators.
Hence let $\Hz_0$ be a Euclidean space (the nuclear part of the rigged 
Hilbert space) with nuclear topology; we put 
$\Hz=C^{\infty }(\Rz,\Hz_0)$. For the standard
position representation and $p_0=i\hbar \partial_t/c$, $q_0=ct$, 
we have
\lbeq{e.ccr}
p_\mu \lp q_\nu = \eta_{\mu\nu} ~~~\mbox{for } \mu,\nu=0,1,\dots,
\eeq
with the metric 
\[
\eta=\Diag(1,-1,\dots,-1).
\]
Restricted to $\Ez_0=\Lin \Hz_0$, this
gives the setting of traditional quantum mechanics.

\bigskip
{\bf Nonrelativistic classical mechanics.}
As discussed, e.g., in {\sc Marsden \& Ratiu} \cite{MarR},
classical physics can be most conveniently described in terms of 
a Poisson manifold $\Omega$. Let $\{ \pdot , \pdot \}$ be the 
associated Poisson bracket on the algebra $\Ez_0:= C^\infty(\Omega)$
of infinitely differentiable complex-valued functions on $\Omega$.
Then $\Ez = C^\infty (\Rz ^2, \Ez _0)$ (defined as in 
{\sc Kriegl \& Michor} \cite{KriM}) is a Poisson algebra with
{\bf classical bracket}
\[
  f \lp g 
  := \frac{\partial f}{\partial t} \frac{\partial g}{\partial E} 
  - \frac{\partial f}{\partial E } \frac{ \partial g}{\partial t}
  + \{ f,g \}
\]
for $f=f(t,E)$, $g=g(t,E)$, and 
Euclidean with {\bf classical integral}
\[
\sint f=\int dt~dE\int_\Omega f(t,E)
\]
(where $\int_\Omega$ is the Liouville measure).
Integrable quantities are the Schwartz functions on $\Ez$.
(Thus integrability in the present sense is much stronger than 
Lebesgue integrability. This is due to our requirement (E1)
which implies that $I\Ez$ must be an ideal in $\Ez$.)
Again, the axioms are easily verified. With $p_0=E/c$, $q_0=ct$
and the standard symplectic Poisson bracket, 
we get again \gzit{e.ccr}.

\section{Physical systems}\label{s.phys}

\hfill\parbox[t]{8.8cm}{\footnotesize 

{\em 
\dots da{\ss} die \"ubliche Quantisierungsvorschrift  
sich durch eine andere Forderung  
ersetzen l\"a{\ss}t [...]\\
Die neue Auffassung ist verallgemei\-nerungs\-f\"ahig und r\"uhrt, 
wie ich glaube, sehr tief an das wahre Wesen der Quantenvorschriften.
}

Erwin Schr\"odinger, 1926 \cite{Sch}\\

}\nopagebreak

Motivated by the prelude, and consistent with the introductory remark 
in the seminal paper ``Quantisierung als Eigenwertproblem'' 
(``Quantization as eigenvalue problem'') 
by {\sc Schr\"odinger} \cite{Sch},
we generalize the Schr\"odinger picture of 
traditional quantum mechanics as follows.
A {\bf physical system} is characterized by a Hermitian {\bf density} 
$\rho \in I\Ez$ with $\rho \ge 0$.
The density, or any set of parameters from which the density can be 
uniquely reconstructed by a well-defined recipe, is referred to as 
the {\bf state} of the system. 
A physical system with density $\rho$ defines {\bf expectations}
\lbeq{e.exp}
  \< f \> := \sint \rho f = \sint f \rho .
\eeq
The {\bf centralizer} $\Ez(S)$ of a quantity $S$ (or a vector of 
Lie commuting quantities) is the set of all quantities Lie commuting 
with (all components of) $S$,
\[
\Ez(S)=\{f\in\Ez \mid S\lp f=0\}.
\]
Clearly, $\Ez(S)$ is again a Poisson algebra. For a quantity  
$f\in\Ez(S)$, the {\bf conditional expectation} at a fixed value $s$ 
of $S$ is defined by
\[
\<f\>_{S=s}=\< f \delta(S-s)\>/\<\delta(S-s)\>,
\]
defined via a limit of integrable functions approaching the delta 
function. For example, if $S$ is Hermitian with real spectrum then 
\[
\<f\>_{S=0}=\lim_{\eps\downto 0}
\<(S-i\eps)^{-1}f(S+i\eps)^{-1}\>/\<(S^2+\eps^2)^{-1}\>.
\]
By construction, conditional expectations always satisfy 
$\<1\>_{S=s}=1$; they satisfy the axioms for an ensemble given in 
{\sc Neumaier} \cite{Neu.ens}.

Dynamical predictions are possible only in a system with 
a well-controlled environment. For a system in interaction with an 
arbitrary environment, the expectation satisfies a dynamics determined
by the {\bf Ehrenfest equations}
\lbeq{e.ehr}
\<\Dc\{f\}\>=0 \forall f\in B\Ez
\eeq
with a {\bf forward derivation} $\Dc$, i.e, a 
continuous linear mapping $\Dc\{\cdot\}:\Ez\to \Ez$ mapping bounded
quantities to bounded quantities and satisfying 
\[
\Dc\{f\}^*=\Dc\{f^*\},~~~
\Dc\{f^*f\} \ge \Dc\{f^*\}f + f^*\Dc\{f\}
\]
for all $f\in \Ez$. Physical systems with the same forward derivation
(but in general different densities) are said to follow the same 
{\bf physical law}. Written in terms of the density, \gzit{e.ehr} 
becomes the {\bf generalized Liouville equation}
\lbeq{e.liou1}
\Dc^*\{\rho\}=0
\eeq
with the {\bf Liouville operator} $\Dc^*$ defined (uniquely by (E4)) by
\[
\sint \Dc^*\{\rho\} f=\sint \rho \Dc\{f\}.
\]
We shall discuss general physical systems and their (dissipative) 
properties elsewhere. 

Here we consider isolated systems only, where the physical law is
characterized by a Hermitian {\bf action} $L\in\Ez$ which determines
the forward derivation.
A physical system with density $\rho$ is called {\bf isolated} 
(and $\rho$ is called a {\bf conservative density}) if 
\[
\<L\>=\sint L\rho=0
\]
and the generalized Liouville equation
\lbeq{e.liou2}
  L \lp \rho =0
\eeq
holds. Since 
\[
\<L\lp f\> = \sint \rho(L \lp f) = \sint (\rho\lp L) f
=- \sint (L \lp \rho)f=0,
\]
expectations in isolated systems satisfy the Ehrenfest equations
\[
  \< L \lp f \> =0 \mbox{~~~for all~} f \in I\Ez.
\]
The axioms for a Euclidean expectation algebra imply that 
$\Dc_\pm\{f\}= \pm L \lp f$ is a forward derivation for both signs;
as discussed elsewhere, this reflects the reversible, conservative 
nature of isolated systems. 

If $\rho$ is a conservative density and $f\in \Ez(L)$ then 
\lbeq{e.newdens}
\rho^f:=f\rho f^*
\eeq
is also a conservative density. Thus a large class of conservative 
densities can be constructed from a single one if some quantities 
$f_l$ in the centralizer $\Ez(L)$ are known, since we may apply 
\gzit{e.newdens} with any polynomial constructed from the $f_l$.
This generalizes the traditional construction of states from the 
vacuum by means of creation operators. It is applicable even where 
-- such as for interacting quantum fields in 4 dimensions -- 
no precise mathematical meaning can be given to the
latter construction.


\section{Hamiltonian systems}\label{s.hamil}

Our axioms cover the traditional physics of Hamiltonian systems.
The action corresponding to an arbitrary time-dependent 
Hamiltonian $H(t)$ is defined as 
\[
L=p_0-H(t),
\]
where $p_0=E$ in the classical case and $p_0 = i \hbar \partial _t$
in the quantum case. In both cases,
\[
  p_0 \lp f=- \dot{f}.
\]
(Strictly speaking, the name 'action' fits tradition only for field 
theories. For multi-particle systems, the above expression for $L$ 
is unrelated to  traditional action principles. 
But applying the same machinery which gives the field equations of 
field theory to this unorthodox action happens to produce the correct 
multi-particle dynamics.)

For conservative quantum systems,
$L \lp \rho =0$ implies for $L=p_0-H$: 
\[
  \dot{\rho} = -p_0\lp \rho=-(L+H)\lp\rho = -H \lp\rho,
\]
and we get the standard {\bf quantum} {\bf Liouville equation}
\lbeq{e.qlio}
  i \hbar \dot{\rho} =[H, \rho ]
\eeq
for a conservative nonrelativistic quantum system with 
Hamiltonian $H$. The Ehrenfest equations reduce to their traditional 
form
\[
i\hbar \frac{d}{dt} \<f\>=\<[f,H]\>,
\]
showing that expectations follow a deterministic law.
For conservative classical systems, exactly the same derivation
applies, and we get the {\bf classical Liouville equation}
\lbeq{e.clio}
  \dot{\rho } = \{ H, \rho \}.
\eeq

\section{Pure states}\label{s.pure}

Pure states are the limiting situation (in a suitable completion of 
the space of integrable quantities) of densities extremal with respect 
to the natural order relation. They are of mathematical interest since 
any density can be written as a convex combination of pure states, and
of physical interest for few-particle systems, where states 
can often be considered as approximately pure. (However, states at 
positive temperature are never pure, and the decomposition into pure 
states is, in the quantum case, not unique. Thus pure states describe
idealized situations only.)

{\bf Pure classical states.}
Here extreme states are distributional limits of densities; the
expectations are algebra homomorphisms into $\Cz$ 
(i.e., characters of the algebra) satisfying $\<L\>=0$.
For nonrelativistic classical physics with phase space variable $z$, 
\[
\<f\>=f(t,E,z),
\]
and the condition $\<L\>=0$ fixes the value of $E$ to $E=H(t,z)$.
Hence we may assume $f$ to be independent of $E$.

Thus pure states of a classical nonrelativistic system
are characterized by a pair $(t,z)$ consisting of a time $t$ and
the phase space location $z$ of the system at this time.
The Ehrenfest equations reduce in the limit of pure states to the 
{\bf Hamiltonian dynamics}
\[
\dot f=\{f,H\}.
\]

{\bf Pure quantum states.}
Extreme states are limiting rank 1 densities 
\[
\rho=\psi\psi^*,~~~\psi\in \Hz^*.
\]
The equation
$ L \lp \rho =0$ implies that $\psi$ is a generalized eigenvector of 
$L$. (See, e.g., {\sc Maurin} \cite{Mau} for a mathematical treatment
in terms of nuclear spaces.)
The condition $\<L\>=0$ then implies that the eigenvalue vanishes. 
Note that, since generalized eigenvectors need not be in $\Hz$,
not all expectations need to exist in a pure state; the latter are 
to be regarded only as idealized limits of physical states. 

Thus pure states of an isolated quantum system are characterized 
by a generalized Schr\"odinger equation 
\lbeq{e.schr}
L\psi=0,~~~\psi\in \Hz^*.
\eeq
We call solutions of \gzit{e.schr} 
{\bf pure conservative quantum states}.

As discussed in the prelude, if the action $L$ is translation 
invariant and $p$ is the generator of the translations,
one can find pure conservative quantum states of definite 4-momentum $k$
by solving the equations.
\lbeq{e.cschro}
p_0\psi=mc\psi,~~~\p\psi=k\psi.
\eeq
In particular, pure states of mass $m$ in a rest 
frame can be found by solving the eigenvalue problem
\[
p_0\psi=mc\psi,~~~\p\psi=0,~~~L\psi=0.
\]
This is a {\bf constrained Schr\"odinger equation}, 
cf. Section \ref{s.cschro} below.

The pure conservative quantum states form a vector space $\Hz^\cons$
on which the centralizer $\Ez(L)$ acts since $f\in \Ez(L)$ and
$\psi\in\Hz^\cons$ imply $Lf\psi=fL\psi=f0=0$. In the quantum case,
$f\in \Ez(L)$ iff $f$ commutes with $L$; thus quantities in $\Ez(L)$
can be found, e.g., by solving the eigenvalue problem for $L$. 
Thus we can create from any particular conservative quantum state
a large class of other conservative quantum states provided we
know enough quantities commuting with $L$.

\section{Classical fields}

We discuss here only boson fields.
By using super Poisson algebras and super versions of all concepts,
fermion fields can be handled in an analogous fashion.

Let $\Hz:=S(\Rz^{1,3})$
be the algebra of infinitely differentiable, fast decaying Schwartz 
functions on Minkowski space $\Rz^{1,3}$, and let $V$ be a  
finite-dimensional symplectic space with symplectic form $\Delta$.
Then the field algebra $\Ez:=C^\infty_\poly(\Hz\otimes V^*)$ of 
infinitely differentiable functions $f$ of the field argument 
$\Phi \in\Hz$ with 
$\partial^nf\in C^\infty(\Hz\otimes V^*,(\Hz\otimes V)^{\otimes n})$
and at most polynomial growth is a Poisson algebra with
\[
f\lp g = \int dx~ \Delta\left(\frac{\partial f}{\partial \Phi(x)},
\frac{\partial g}{\partial \Phi(x)}\right).
\]
With expolynomial functions (linear combinations of products 
of polynomials with the exponential of a negative definite, 
quadratic polynomial) as integrable functions, $\Ez$ is Euclidean with 
an integral definable via infinite-dimensional Gaussian measures.

{\bf Pure classical field states.}
A pure state over the field algebra $\Ez=C^\infty_\poly(\Hz\otimes V^*)$
assigns to each $f\in\Ez$ the value $f(\Phi)$ at a particular 
field $\Phi\in \Hz\otimes V^* $.
The Ehrenfest equations reduce in the limit of pure 
states over the field algebra to the equations
\[
L\lp f=0 \forall f\in\Ez.
\]
Inserting the linear function $f=a(\Phi)$, where
\[
a(\Phi):=\D\int dx\, a(x)^T\Phi(x),
\]
into $L\lp f=0$ we get 
\[
\Delta\left(\frac{\partial L}{\partial \Phi(x)},
a(x)\right)=0
\] 
for $a \in \Hz\otimes V$ with compact support. Since $\Delta$ is 
nondegenerate, we conclude
\[
\frac{\partial L}{\partial \Phi(x)}=0 \forall x\in\Rz^{1,3}.
\]
This is the traditional {\bf stationary action principle}.
In the current setting, it is not a 
postulate but a consequence of the Ehrenfest equations.
(The equations for other choices of $f$ are consequences of this.)

\bigskip
To get the traditional field theories, we simply need to find the right 
symplectic structure for each type of field. The field components must 
appear in conjugate pairs, which we arrange to two conjugate vectors 
$\Phi$ and $\Phi^c$ (in place of the single $\Phi$ used before).
Then adequate commutation relations are
\[
a(\Phi)\lp b(\Phi)=a(\Phi^c)\lp b(\Phi^c)=0,
\]
\[
a(\Phi^c)\lp b(\Phi)=(a|b):=\int dx\, a(x)^Tb(x),
\]
where $\Phi^c=\Phi^*$ for complex fields (which come in complex 
conjugate pairs), while for real fields $\Phi$ and $\Phi^c$ are
independent. 
For real fields which have no conjugate partner in the
Lagrangian, one adds additional conjugate partners to the algebra
of quantities. These additional fields are -- like gauge degrees 
of freedom -- unobservable and do not affect the field equations 
for the original fields.

Hence the present framework allows a consistent implementation 
of all classical field equations derivable from the stationary 
action principle.
({\em Note:} If we apply this to the 
electromagnetic 4-vector potential, we get, in contrast to the 
approach in canonical quantization, a conjugate 4-vector potential,
with standard symplectic Lie bracket for each component!)

By extending the above framework to Euclidean super Poisson
algebras, one can also incorporate classical fermion fields. 
In particular, we can implement a 
{\bf classical} version of the {\bf standard model}, 
{\bf including gravitation} within the
present setting.

If we use in place of symplectic Poisson algebras suitable 
Lie-Poison algebras, the Ehrenfest equations produce in the limit of 
pure classical states for appropriate actions both the relativistic 
\cite{MisTW} and nonrelativistic \cite{Mor} {\bf Euler equations} 
for perfect fluids and the {\bf Euler-Poincar\'e equations} \cite{MarR}.
Using suitable Lie-Poison algebras of functions of phase space 
fields, it is possible to define natural actions for which the
Ehrenfest equations produce in this way the Vlasov equations. 
In suitable tensor products one can then form actions that define
Vlasov equations interacting with electromagnetic and/or gravitational 
fields, giving {\bf Vlasov-Maxwell equations} (cf., e.g., \cite{Rei})
and {\bf Vlasov-Einstein equations} (cf., e.g., \cite{And}). 

Details will be given elsewhere.

\section{Phase space quantization}

There are many ways to quantize a classical system. From the point of 
view of being able to do analysis (i.e., error estimates), the 
mathematically most developed form is deformation quantization 
(see, e.g., {\sc Rieffel} \cite{Rie}), which deforms a commutative 
product into a Moyal product. 
In the following, we propose an alternative deformation approach which,
instead, {\bf deforms the operators} $f \in \Ez$ by  
embedding $\Ez$ into $\Lin \Ez$, identifying  
$f \in \Ez$ with the multiplication mapping  
$g \to fg$. This can be done with surprising ease.

The superoperators $M_f$ and $D_f$ defined by
\[
  M_f\{g\}:=fg,~~~D_f\{g\}:=f \lp g
\]
belongs to $\Lin \Ez$.
For $f\in\Ez$, we define the {\bf quantization} $\widehat{f}$ of $f$ by
\[
  \widehat{f} :=M_f-\frac{i \hbar}{2} D_f \in\Lin\Ez.
\]
The expectations
\[
  \< \widehat f \> = \< f \> - \frac{i \hbar }{2} \< D_f \> 
\]
differ from those of $f$ by a term of order $O(\hbar)$, justifying an 
interpretation in terms of ``deformation''. In particular, we
automatically have a good classical limit.
 
To actually quantize a classical theory, one may choose a 
Lie algebra of relevant quantities generating the Poisson algebra, 
quantizes its elements by the above rule, expresses the classical
action as a suitably ordered polynomial expression in the 
generators, and uses as quantum action this expression with all
generators replaced by their quantizations.

In general, the above recipe for phase space quantization gives an 
approximate Poisson isomorphism, up to $O(\hbar)$ terms. 
But Lie subalgebras are mapped into (perhaps slightly bigger)
Lie algebras, and one gets a true isomorphism for all embedded 
Heisenberg Lie algebras, i.e., Lie algebras 
where all Lie products are multiples of a central element $1$.

{\bf Quantization Theorem.}  
If $\Ez$ is commutative then the quantum bracket 
\[
A\lp B = \iota[A,B]~~~\mbox{for } A,B\in\Lin \Ez
\]
satisfies, for $f,g\in\Ez$,
\[
  \widehat{f}\lp\widehat{g}=M_{f \slp g} - \frac{i\hbar}{4} D_{f\slp g}
  = \half (M_{f \slp g} + \widehat{f \lp g}),
\]
Any Lie subalgebra $\Lz$ of $\Ez$ defines a Lie algebra 
\[
  \widehat{\Lz}= \{ M_{f \slp g} + \widehat{h} \mid f,g,h \in \Lz \}
\]
under the quantum bracket. If $\Lz$ is a Heisenberg Lie algebra then 
$\widehat{~}:\Lz\to\widehat\Lz$ is a Lie isomorphism.

The proof is not difficult but will be given elsewhere.

In particular, for the standard symplectic Poisson algebra 
$\Ez=C^\infty (\Rz^n \times \Rz^n)$, phase space quantization
amounts to using the reducible representation
\[
 \widehat p = p- \frac{i \hbar }{2} \partial_q,~~~
 \widehat q =q+ \frac{i \hbar }{2} \partial_p
\]
of the canonical commutation rules on phase space functions
instead of the traditional irreducible representation
\[
  \tilde p=-i \hbar \partial_x,~~~\tilde q=x
\]
on configuration space functions.
It will be shown elsewhere that these representations are related by a 
{\bf Wigner transform} (cf. {\sc Wigner} \cite{Wig}).

By quantizing in phase space, one gives up irreducibility
(and hence the description of a state by a {\em unique} density)
but gains in simplicity. Perhaps this is comparable to the situation
in gauge theory, where the description by gauge potentials introduces
some arbitrariness with which one pays for the more elegant 
formulation of the field equations but which does not affect the
observable consequences.

\section{Quantum field theory}\label{s.qfield}

   \hfill\parbox[t]{8.4cm}{\footnotesize 

{\em
A good many physicists are now working on the problem of trying to 
set up a quantum field theory independently of any Hamiltonian.
[...]\\
I still think that in any future quantum theory there will have to be 
something corresponding to Hamiltonian theory, even if it is not in the 
same form as at present.
}

Paul Dirac, 1964 \cite{Dir}

}\nopagebreak

Actions for classical or quantum field theories are based on 
representations of a symmetry group and corresponding invariant actions.
In any fundamental theory, the symmetry group must contain either the 
Galilei group (for nonrelativistic fields) 
or the Poincar\'e group (for relativistic fields); if gravitation is 
involved, the symmetry group must also contain the group of all 
diffeomorphisms of some spacetime manifold.

Having a symmetry group is equivalent with having nonuniqueness in
the description of a physical system. Different states providing 
equivalent descriptions (satisfying the same laws but in different 
coordinates) are commonly said to correspond to different choices 
of an inertial system. 
Changing the inertial system used to coordinatize a system 
changes the state and hence the expectations; for example, moving an
inertial system $O$ (illustrated by an observing intelligent robot) 
in time produces a change in observed expectations which $O$ conceives 
of as the intrinsic dynamics of the environment, 
while moving (rotating or translating) the inertial system $O$ in 
space produces a change in observed expectations which $O$ conceives 
of as the illusion of the space moving around it caused by the motion
of its moving head. We now formalize these considerations.

Let $\Lz$ be the Lie algebra of the Galilei group, the Poincar\'e 
group or any assumed symmetry group containing one of these groups, 
with Lie product $\lp$. 
Let $p\in \Lz^{1,3}$ be the generator 
of the translation subgroup in the canonical basis. 
Let $J$ be a Poisson representation of $\Lz$ in a Euclidean Poisson 
algebra $\Ez$, defined by
\lbeq{e.wig1}
J(\delta)\lp J(\delta')=J(\delta \lp \delta')
\forall \delta, \delta'\in\Lz.
\eeq
$P:=J(p)$ (taken componentwise) defines the (total) 
{\bf physical 4-momentum}.
A smooth change of the inertial system 
(modeling a virtual motion of the robots head) 
is described by an arbitrary continuously differentiable mapping 
$\delta:[0,1]\to \Lz$ specifying the infinitesimal motions 
$\delta(\tau)\in\Lz$ of the inertial systems at 
instant $\tau\in[0,1]$. A corresponding assignment of densities 
$\rho(\tau)\in I\Ez$ at instant $\tau$ is called {\bf consistent} 
if it satisfies the differential equation
\lbeq{e.change}
\frac{d}{d\tau}\rho(\tau)=\rho(\tau) \lp J(\delta(\tau)).
\eeq
In classical physics, this describes a canonical,
in quantum physics a unitary transformation representing a general
element of the (connected part of the) symmetry group.
In particular, an observer moving in space-time with uniform 
velocity $u\in \Rz^{1,3}$ finds the density changing according to the 
{\bf covariant Liouville equation}
\lbeq{e.covsch}
\frac{d}{d\tau}\rho(\tau)=\rho(\tau) \lp J(u\cdot p).
\eeq
Thus we have a covariant generalization of the nonrelativistic
situation considered in Section \ref{s.hamil}.

Since such a change of inertial systems should not affect the physics,
we require that quantities (and in particular the action $L$, i.e., 
the physical law) are unaffected by these changes, and that an 
isolated system remains isolated. The former condition is simply
the requirement that we base our setting on the 
{\bf Schr\"odinger picture}, and the latter condition amounts to
\lbeq{e.inv}
L\lp \rho(\tau)=0 \forall \tau
\eeq
whenever $L\lp \rho(0)=0$. To analyze this condition, let $\rho(0)$
be the density of an isolated system, and put
\[
e(\tau):=L\lp \rho(\tau).
\]
Then $e(0)=0$ and
\[
\bary{lll}
\D\frac{d}{d\tau}e(\tau)
&=&\D\frac{d}{d\tau}(L\lp \rho(\tau))=L\lp \frac{d}{d\tau}\rho(\tau)
=L\lp (\rho(\tau) \lp J(\delta(\tau)))\\~\\
&=&(L\lp \rho(\tau))\lp J(\delta(\tau))
+\rho(\tau) \lp (L \lp J(\delta(\tau)))
\eary
\]
so that
\lbeq{e.ode}
\frac{d}{d\tau}e(\tau)=
e(\tau)\lp J(\delta(\tau))+ \rho(\tau) \lp (L \lp J(\delta(\tau))).
\eeq
If \gzit{e.inv} holds then $e(\tau)$ vanishes identically, and
this reduces to $\rho(\tau) \lp (L \lp J(\delta(\tau)))=0$. 
The requirement that this holds for arbitrary densities and arbitrary
smooth changes of the inertial system therefore demands that
\lbeq{e.wig2}
L\lp J(\delta) \in C\Ez \forall \delta\in\Lz,
\eeq
where $C\Ez$ denotes the {\bf Lie center} of $\Ez$, the algebra of
quantities which Lie commute with all quantities. An action $L$
satisfying \gzit{e.wig2} is called {\bf $\Lz$-invariant}. Conversely,
if the action $L$ is $\Lz$-invariant, then \gzit{e.ode} reduces to
\[
\frac{d}{d\tau}e(\tau)=e(\tau)\lp J(\delta(\tau)).
\]
Under conditions which guarantee the unique solvability of the
initial value problem \gzit{e.ode}, we conclude that 
$e(\tau)=0$ for all $\tau$, proving \gzit{e.inv}.
Thus the $\Lz$-invariance of the action is essentially equivalent to
the requirement that being isolated is a covariant concept.

In particular, using in our setting a Poincar\'e invariant action $L$
defines a relativistic physical theory. 
As shown in Sections \ref{s.phys} and 
\ref{s.pure}, we can use an arbitrary conservative density (resp.\ pure 
quantum state) and a set of quantities in the centralizer $\Ez(L)$ to 
construct a large class of conservative densities (resp.\ pure quantum 
states) as possible states of an isolated physical system with given
action. 

\bigskip
Having phase space quantization as a universal generalization of the 
Wigner transform, we can use it to quantize the basic fields of any 
(Galilei or Poincar\'e invariant) classical field theory. This gives 
well-defined mathematical definitions of the various (nonrelativistic
or relativistic) quantum field theories in current physical usage.

Using a Galilei invariant action one gets nonrelativistic 
field theory. As explained nonrigorously in many textbooks (e.g., 
{\sc Umezawa} et al. \cite{UmeMT}), nonrelativistic quantum field 
theory is in principle equivalent to nonrelativistic quantum mechanics. 
Therefore, one uses for nonrelativistic problems field theory only 
to describe bulk matter, while scattering and bound state problems 
are handled with the Schr\"odinger equation. This is much simpler 
than solving the full operator dynamics of field theory. 

In relativistic quantum field theory, there has been in the past
no analogue of the Schr\"odinger equation that could have been used
for this purpose. Thus even simple scattering problems were formulated 
in a field theoretic language accessible to a perturbative treatment, 
and bound state problems (see, e.g., {\sc Weinberg} \cite{Wei}) 
could be described only very indirectly through poles in the S-matrix.
For the latter, there is no sound mathematical basis since in 
traditional quantum field theory, the S-matrix is only defined 
perturbatively in terms of a presumably divergent 
({\sc Dyson} \cite{Dys2}) asymptotic expansion,
so that, mathematically, talking about its poles is nonsense.

The results of the present paper show, however, that to each quantum 
(field or particle) theory there is a corresponding 
constrained Schr\"odinger equation from which one can construct
pure conservative quantum states with definite momentum in complete 
analogy to the nonrelativistic case,
and without restriction to a particular symmetry group.
(Mathematically, it is suspect if certain techniques work for
a particular, highly nontrivial group but not for all groups. Already
from this perspective one could see that something was missing from 
current quantum field theory!)

\section{Wightman axioms}

   \hfill\parbox[t]{10.5cm}{\footnotesize 

{\em
The quantum theory of fields never reached a stage where one could 
say with confidence that it was free from internal contradictions -- 
nor the converse. In fact, the Main Problem [...] turned out to be 
[...] to show that the idealizations involved in the fundamental 
notions of the theory are incompatible in some physical sense, 
or to recast the theory in such a form that it provides a practical 
language for the description of elementary particle dynamics.
}

R.F. Streater and A.S. Wightman, 1963 \cite{StrW}

}\nopagebreak

Traditionally, mathematical physicists approach relativistic quantum 
field theory via an axiomatic approach discussed in detail by 
{\sc Streater \& Wightman} \cite{StrW}. The Wightman axioms 
({\sc Wightman} \cite{Wigh}) are an interpretation of field theory 
not in terms of field equations but in terms of correlation functions. 
Relations to the Lagrangian approach have been lacking so far.
But one would have such relations if one could combine tradition with
the present formulation of quantum field theory.
Thus one would like to realize the Wightman axioms by identifying the 
vacuum with a pure conservative quantum state $\psi_0$ with zero 
momentum, i.e.,
\lbeq{e.wig3}
L\psi_0=0,~~~ P\psi_0=0,
\eeq
and (in view of the remarks at the end of Section \ref{s.pure})
Wightman field operators by suitable Hermitian quantities 
in the centralizer $\Ez(L)$. 

It is not clear whether the Wightman axioms describe correctly the 
structure of relativistic quantum states. Apart from generalized free 
fields, no realization of the Wightman axioms in 4-dimensional 
space-time is known  (see, e.g., {\sc Rehren} \cite{Reh}), 
and there are no-go theorems -- stating, for example, that
there is no natural interaction picture \cite[Theorem 4-16]{StrW} --
pointing to the possibility 
that these axioms are indeed too strong to describe realistic theories. 

To prove that the assumptions defining a Wightman field can (or cannot)
be satisfied in the present context is therefore a highly nontrivial 
task. But at least it is embedded into a well-defined functional 
analytic context, where the Poincar\'e representation is already fixed.
This might make it tractable for systems like QED, which are close 
to nonrelativistic quantum mechanics. 
Therefore, one might be able to adapt the insights from 
nonrelativistic scattering theory (which provides a diagonalization of
the action and hence full control over its centralizer) to the new 
situation.

On the other hand, even without knowing the existence of Wightman 
fields (and even if one could prove that they do not exist), 
the setting presented here makes sense and defines for arbitrary 
actions a good quantum field theory, closely related to physical 
practice. In particular, one can try to generalize to the new 
constrained Schr\"odinger equations the supply of techniques available 
for ordinary Schr\"odinger equations, and in this way complement the 
current perturbative techniques of quantum field theory by 
techniques known from nonrelativistic quantum mechanics.
A first step in this direction -- the generalization of the 
projection formalism -- has been done already; 
see {\sc Neumaier} \cite{Neu.proj}. Work on scattering theory is 
under way.

\section{Phenomenological relativistic dynamics}

   \hfill\parbox[t]{10.6cm}{\footnotesize 

{\em
In spite of the acceptance of field theories as a matter of principle, 
most realistic dynamical calculations in nuclear physics, and many 
in particle physics, utilize the nonrelativistic Schr\"odinger 
equation. [...] 
Relativistic direct interaction theories of particles lie between
local field theoretical models and nonrelativistic quantum mechanical 
models.
}

B.D. Keister and W.N. Polyzou, 1991 \cite{KeiP}\\

}\nopagebreak

While fields are usually used to describe nature on a fundamental 
level, practical work (especially for bound states and resonances)
requires phenomenological few-particle equations, which are frequently
related only loosely to underlying fields; see the references in the
next section. It is therefore interesting
to see that a variety of covariant phenomenological few-particle 
equations can be easily built in the present framework.
We do this by using Poincar\'e invariant actions on Hilbert spaces 
carrying a suitable Poincar\'e representation without states of 
negative energy. 

The possible irreducible Poincar\'e representations (modeling 
elementary particles) were classified by {\sc Wigner} \cite{Wig2}. 
The representations of positive (relativistic) energy
take their simplest form in momentum space; the momenta $p$ are 
restricted to a mass shell 
\lbeq{e.shell}
\Omega(\tilde p)=\{p\in\Rz^{1,3} \mid p^2=\tilde p^2,~ p_0>0\}, 
\eeq
the orbit of a 4-vector $\tilde p$ under the Poincar\'e group.
It is possible to combine these irreducible Poincar\'e representations
in many ways to obtain reducible momentum space representations for 
few-particle systems. Traditionally (see, e.g., 
{\sc Weinberg} \cite{Wei1} for the canonical field quantization 
approach and the review in {\sc Keister \& Polyzou} \cite{KeiP} for 
the direct relativistic Hamiltonian few-body approach), this is done by 
breaking the manifest invariance to a maximal subgroup of the 
Poincar\'e group, with all the awkwardness this entails. 

The key that allows us to preserve a manifestly covariant formalism, 
thus overcoming the traditional problems in canonical quantization, 
is the fact that we use as algebra of quantities the linear operators 
on a space of wave functions slightly bigger than traditional Fock 
spaces. This is done in the following by adding a velocity vector
$u$ as a dynamical parameter, which allows us to deform the
bare mass shell $p^2=(mc)^2$ (where $c$ is the speed of light)
to $p^2=(mu)^2$, which in turn permits the
conservation of total 4-momentum in interactions.

A phenomenological realization of a system of $N$ massive scalar 
particles with {\bf rest masses} $m^1,\dots,m^N>0$ and {\bf charges} 
$Q^1,\dots,Q^N$ is now realized by wave functions 
\[
\psi=\psi(u,p^{1:N})=\psi(u,p^1,\dots,p^N)
\]
whose coordinates are a global {\bf 4-velocity} vector $u$ with 
$0<u\in\Rz^{1,3}$ and the particle {\bf 4-momentum} vectors $p^a$ in 
the {\bf dynamic mass shells} $\Omega(m^au)$ whose scale depends on $u$.
The {\bf total 4-momentum} $\sum p^a$ is required to be parallel to 
the 4-velocity $u$. Thus the space of wave functions is
\lbeq{e.3Nplus1}
\Hz=C^\infty(\Omega^N),
\eeq
where $\Omega^N$ is the set of all tuples
\[
(u,p^{1:N})=(u,p^1,\dots,p^N)
\]
with 
\[
p^a\in\Omega(m^au) \mbox{~~for } a=1,\dots,N,~~~~~
0< u~\parallel\sum p^a.
\]
The (not everywhere defined) Hermitian inner product 
-- from which a Hilbert space can be constructed by completing the 
space of vectors of finite norm -- is given by
\[
\phi^*\psi:=\int dm~du Dp^1\dots Dp^N \delta\Big(mu-\sum p^a\Big)
\overline{\phi(u,p^{1:N})}\psi(u,p^{1:N}),
\]
where 
\lbeq{e.covint}
Dp=dp~\delta(p^2-(mu)^2)
=\frac{d\p}{2p_0}=\frac{d\p}{2\sqrt{(mu)^2+\p^2}}
\eeq
is the invariant measure on a dynamic mass shell $\Omega(mu)$. 
The {\bf one-particle operators} are defined as
\[
J(f):=\sum_a f(u,Q^a,m^a,p^a,M^a),
\]
where the diagonal operator $f=f(u,Q,m,p,M)$ is a function of 
4-velocity $u$, charge $Q$, mass $m$, 4-momentum $p$ and 
{\bf 4-angular momentum} 
\lbeq{e.M}
M:= p \wedge \frac{\partial}{\partial p}
\eeq
with components
\[
M_{\mu\nu}=p_\mu \frac{\partial}{\partial p_\nu}
-p_\nu \frac{\partial}{\partial p_\mu},
\]
and the superscript $a$ indicates application to the coordinates of the
$a$th particle. (Note that the global 4-velocity $u$ carries no
superscript; it is shared by all particles.) 
Since the $p_\mu$ are the Poincar\'e generators of 
translation in the direction of the $\mu$-axis and the $M_{\mu\nu}$
are the standard generators of the Lorentz transformations,
it is easy to see that the total 4-momentum $J(p)$ and 
{\bf total 4-angular momentum} $J(M)$ define a representation of the 
Poincar\'e group {\em without negative energy states}.
In the terminology of {\sc Dirac} \cite{Dir.point}, 
it is a representation in the point form. (It shares this property with
the representations of {\sc Ruijgrok} \cite{Rui} which are based on
Lippmann-Schwinger equations. But his translation generators are much 
more complicated than the present ones.)

On the space \gzit{e.3Nplus1}, one can now define actions of the form
\lbeq{e.phenaction}
L=L_0- V,
\eeq
where the {\bf kinetic action} $L_0$ is a Poincar\'e invariant 
one-particle operator,  and the {\bf interaction} $V$ is a 
Poincar\'e invariant integral operator.

\section{Poincar\'e invariant multiparticle interactions}

For scalar particles, the simplest covariant kinetic action is
\lbeq{e.Lboson}
L_0=J\Big(\frac{p^2-(mc)^2}{2m}\Big)=J\Big(\frac{m}{2}(u^2-c^2)\Big),
\eeq
with a constant $c>0$, the {\bf speed of light}. However, more 
complicated covariant formulas with rational or analytic dependence 
on $m$ and $p^2$ are admissible, too, if they vanish for $p^2=(mc)^2$
and nowhere else. In this case, the generalized 
Schr\"odinger equation $L\psi=0$ implies for noninteracting particles,
where $V=0$, the relation $u^2=c^2$, forcing the
dynamic mass shells to equal the bare mass shells.
 
To construct a versatile class of Poincar\'e invariant interactions, 
we first note that the vector
\lbeq{e.massproj}
p_m:=p+u\frac{-p\cdot u+\sqrt{(p\cdot u)^2-p^2u^2+(mu^2)^2}}{u^2}
\eeq
is in the dynamic mass shell $\Omega(mu)$. Indeed, it suffices 
by covariance to check the case where $\uu=0$; then $u_0>0$, 
$u^2=u_0^2$, $p\cdot u=p_0u_0$,
\[
(p_m)_0=p_0+u_0\frac{-p_0u_0+\sqrt{\p^2u_0^2+(mu_0^2)^2}}{u_0^2}
=\sqrt{\p^2+(mu_0)^2}>0,
\]
and since $\p_m=\p$, we find $p_m^2=(mu_0)^2=(mu)^2$.
Thus the mapping $p\to p_m$ (the dependence on $u$ is not written 
explicitly) is a nonlinear projection to the dynamic mass shell
$\Omega(mu)$. 

The simplest choice for a nontrivial interaction is a sum of pair 
interactions,
\lbeq{e.pairint}
V=\sum_{a<b} V^{ab},
\eeq
where $V^{ab}=V^{ba}$ acts on the coordinates of particles $a$ and $b$ 
as
\lbeq{e.inter}
(V^{ab}\psi)(u,p^a,p^b)=\D\int dq~\delta(u\cdot q)
U^{ab}(q)\psi(u,(p^a+q)_{m^a},(p^b-q)_{m^b}),
\eeq
where the projections are to be taken with respect to the common
4-velocity argument $u$, and $U^{ab}(q)$ is also allowed to depend 
on mass, momentum and charge of the particles $a$ and $b$.
The delta function removes a redundancy in the projections, 
which do no change if a multiple of $u$ is added to $q$. 
The construction is such that $V^{ab}$ is automatically translation 
invariant. In particular, if all $U^{ab}$ are 
Hermitian and Lorentz invariant then $V$ and hence the action
\gzit{e.phenaction} is Hermitian and Poincar\'e invariant. 
For example, this is the case in pair potentials of the form
\lbeq{e.relcoul}
 U^{ab}(q)=\re~ \frac{
\beta_S (m^am^b)^2
+(\beta_V m^am^b+\alpha Q^aQ^b)p^a\cdot p^b
+\beta_T (p^a\cdot p^b)^2
}{m^am^b(q^2+i\eps)},
\eeq  
where the limit $\eps\downto 0$ is to be taken in \gzit{e.inter} to
regularize the potential near $q=0$. These potentials describe 
relativistic electromagnetic and gravitational forces;
the coupling constants $\alpha$ and $\beta_S,\beta_V,\beta_T$ 
determine the strength of the {\bf electromagnetic} and the scalar, 
vector, and tensor {\bf gravitational interaction}, respectively.
(This will be justified in the next section by considering the 
nonrelativistic limit.)
By making these coupling constants $q$-dependent 
({\bf running coupling constants}), one can also account covariantly 
for phenomenological self-energy contributions; cf. the discussion in
{\sc Peskin \& Schroeder} \cite[pp. 252--255]{PesS}.

Note that after Fourier transform into spacetime, we get -- 
in contrast to field theories -- a nonlocal (but still Poincar\'e 
invariant) action.

\bigskip
This basic setting can be extended in various ways.
Particles with positive spin or with internal symmetries are easily 
accommodated, especially when using the representations 
discussed in {\sc Weinberg} \cite{Wei1,Wei2}. (They are of course
equivalent to Wigner's representations but computationally more 
tractable.) Particles with positive integral spin
are handled in exactly the same way, except that the wave functions 
have additional indices, the angular momentum gets an additional 
intrinsic spin term operating on these indices, and the inner product 
has a slightly different form. It is easy to specify  $\Lz$-invariant 
interaction terms similar to \gzit{e.relcoul} for particles with 
positive spin and for particles with inner symmetries 
(and corresponding matrix-valued charges $Q^a$); but such interactions 
are now also restricted by Clebsch-Gordan rules (cf. 
{\sc Weinberg} \cite{Wei3}). 

Fermion particles with half-integral spin are handled similarly, using 
spinor components in the wave functions and kinetic actions such as
\lbeq{e.Lfermion}
L_0=J(p\cdot \gamma c-mc^2).
\eeq
The resulting constrained Schr\"odinger equations
\[
J(p\cdot \gamma c-mc^2)\psi = V\psi,~~~ p\psi=k\psi
\]
generalize the Dirac equations to the multiparticle case.
Details about the handling of spin will be given elsewhere.

Massless particles are handled in the same way, except that the kinetic 
part of the action is absent, since these particles never go off-shell 
in our phenomenological setting.

For indistinguishable particles, symmetrization and antisymmetrization 
can be done in the standard way. Different kinds of particles are 
handled by adding to the sum of their self-actions another interaction.
Few-particle systems in which the particle number is not conserved
can be modelled by using a direct sum of Hilbert spaces of the type 
\gzit{e.3Nplus1} and covariant interactions changing the particle 
number.
For example, we may model the emission and absorption of a photon 
of momentum $p$ by a massive scalar particle of charge $Q^a$
with the Hermitian and Poincar\'e invariant interaction proportional to
\[
(V\psi)(u,p,p^a)= F(p)\frac{Q^ap^a}{m^a} \psi(u,(p^a+p)_{m^a}),
\]
\[
(V\psi)(u,p^a)=\D\int dp~
\delta(p^2)F(p)\frac{Q^ap^a}{m^a}\cdot\psi(u,p,(p^a-p)_{m^a}),
\]
where the {\bf form factor} $F(p)$ is an arbitrary 
covariant scalar $C^\infty$-function formed from $p$, $p^a$ and $u$.
(Note that the photon wave function has additional vector components, 
with respect to which the inner product $\cdot$ is taken.) 
Previous covariant few-particle models could not handle this situation 
({\sc Keister \& Polyzou} \cite[p. 392]{KeiP}).

In a multiparticle system, 
one can model in the same way the interactions corresponding to 
Feynman diagrams with a single vertex of degree 3, 
and in a similar way also interactions corresponding to more 
complex vertices. Note that because momentum is conserved and all 
particle energies are positive, particles cannot be created from a 
vacuum state (with $0$ particles), nor can particles be annihilated 
without creating (or preserving) at least one particle. 
Thus the phenomenological approach does not have the problems which
field theories have with the presence of an interacting (`fluctuating') 
vacuum.

We see that the possibilities for the new action-based relativistic 
models fully match (and even exceed) the freedom available for 
nonrelativistic Hamiltonian systems. 
Since they are manifestly Poincar\'e invariant, 
they are much simpler than various relativistic
Hamiltonian models that have been constructed in the past (see, e.g., 
the review in {\sc Keister \& Polyzou} \cite{KeiP} for nuclear
physics, {\sc Crater} et al. \cite{CraBW} for QED, and {\sc Ruijgrok}
\cite{Rui} for a Lippmann-Schwinger based model), but have the 
same advantages as the latter: consistency with relativity theory, 
tractable few-body calculations, easy treatment of bound states,
resonances, and particle production, 
and easy fit to parametric models. In addition, 
they can be used to give phenomenological models of quantum systems 
in which the particle number is not preserved, or the spin is $>1$.

In time, such action-based relativistic models may therefore replace 
the many 
nonrelativistic (e.g., {\sc Isgur} \cite{Isg}, {\sc Karl} \cite{Kar}),
semirelativistic (e.g., {\sc Lucha} et al. \cite{LucS,LucSG}) and 
relativistic (e.g., {\sc Keister \& Polyzou} \cite{KeiP})
Hamiltonian approximations, and approximations based on Bethe-Salpeter  
equations (e.g., {\sc Kummer \& M\"odrich} \cite{KumM}) or 
Dyson-Schwinger equations (e.g., {\sc Roberts \& Williams} \cite{RobW})
now in vogue for the phenomenological description of quarks, mesons, 
baryons, and other relativistic matter. 
Since our phenomenological actions are easily 
made manifestly symmetric under the full symmetry group of a system, 
it may also give more workable low energy effective theories for the 
standard model, such as chiral perturbation theory 
(e.g., {\sc Ecker} \cite{Eck,Eck2}) or quantum hadrodynamics 
(e.g., {\sc Serot} \cite{Ser}, {\sc Serot \& Walecka} \cite{Ser,SerW}).

The relation between the above action-based relativistic 
multiparticle models and
the field-theoretic models discussed earlier is not clear at present.
It is expected that the projection techniques from  
{\sc Neumaier} \cite{Neu.proj} relate the field theories from 
Section \ref{s.qfield} to corresponding effective $N$-particle theories 
modeled as in the present section. On the other hand, it is also 
conceivable that the field theories should rather be regarded as 
limits of $N$-particle theories in the thermodynamic limit
$N\to\infty$. There are indications that this might be the case for 
QED (since radiation phenomena are always dissipative) and for 
gravitation (since black hole thermodynamics changes pure states to 
mixed states, cf. {\sc Wald} \cite[pp. 180--185]{Wal}; the traditional 
coupling to a hydrodynamic model is also meaningful only in a 
thermodynamic limit).

\section{Constrained Schr\"odinger equations}\label{s.cschro}

States of fixed total 4-momentum $J(p)$ can be obtained by solving 
\gzit{e.cschro}. With a Lorentz boost, we may transform
the system to a rest frame; the resulting constraint $J(\p)=0$ can 
be imposed kinematically by restricting the 4-velocity to $\uu=0$.
Since $c$ and $J(mc^2)$ are constants, the wave function is an 
eigenstate of the {\bf rest frame energy} $J(p_0c-mc^2)$ 
(a shifted relativistic energy $p_0c$, introduced in analogy to the 
prelude), and we are left with the (still rotation invariant) 
{\bf constrained Schr\"odinger equations} 
\lbeq{e.cschro2}
\psi=\delta(\uu)\psi_0,~~~L\psi=0,~~~J(p_0c-mc^2)\psi=E\psi,
\eeq
the relativistic analogue of the nonrelativistic multiparticle
Schr\"odinger equation after separation of the motion of the center 
of mass. Thus our phenomenological approach is a covariant 
version of the situation in the prelude: The mass shells form 
3-dimensional manifolds, and the momenta $p^a$ can be considered 
as relativistic analogues of 3-momentum vectors. Since $\uu=0$, 
the 4-velocity contributes only one 
additional degree of freedom $u_0$, which replaces the energy degree 
of freedom of the nonrelativistic situation. Thus, in contrast to the 
realizations of quantum field theory discussed above, to 
traditional Bethe-Salpeter equations, and to proper time based 
relativistic multiparticle dynamics (see, e.g., 
{\sc Fanchi} \cite{Fan}), there are no superfluous 
degrees of freedom, but the treatment is still manifestly covariant.

The delta function in the interaction \gzit{e.inter} forces $q_0=0$. 
Dropping the redundant coordinates $\uu=0$, 
$p_0=\sqrt{(mU_0)^2+\p^2}$ and $q_0=0$ from the notation,
the interaction can be written as the 3-dimensional integral
\lbeq{e.inter.nr}
(V^{ab}\psi)(u_0,p^a,p^b)= 
c^{-1}\int d\q~ U^{ab}(\q)\psi(u_0,\p^a+\q,\p^b-\q);
\eeq
the prefactor comes from the delta function in \gzit{e.inter}.
If we now Fourier transform in space to get the position representation,
\[
\widehat\psi(u_0,\x^a,\x^b)=\int d\p^ad\p^b
e^{\iota\p^a\cdot\x^a}e^{\iota\p^b\cdot\x^b}\psi(u_0,\p^a,\p^b),
\]
we find 
\[
\widehat{V^{ab}\psi}(u_0,\x^a,\x^b)=
\widehat{U^{ab}}(\x^b-\x^a)\widehat \psi(u_0,\x^a,\x^b)
\]
with the spatial potential
\lbeq{e.Vpos}
\widehat{U^{ab}}(\r)=c^{-1}\int d\q~e^{\iota\q\cdot\r}U^{ab}(\q).
\eeq
This looks like a nonrelativistic formula, but the covariant nature
of the model is visible in the form \gzit{e.inter} of $U^{ab}(q)$
and also shows in the constraint nature of \gzit{e.cschro2}.
Compared to the nonrelativistic case, this is now a general linear 
eigenvalue problem for the eigenvalue $E$, and its solution is 
slightly more demanding. But numerical methods are available; see,
e.g., {\sc Golub \& van Loan} \cite{GolvL}.

\bigskip
{\bf The nonrelativistic limit.}
To deepen the analogy, we give a rough, heuristic derivation of the
nonrelativistic limit $c\to\infty$; it would be interesting to have a
rigorous version of this from which one can obtain error bounds.
The equation $L\psi=0$ can be written (for bosons) as
$J(\frac{1}{2} m (u^2-c^2))\psi=V\psi$. For small potential energies, 
$V\ll J(m)c^2$ and small spatial momenta, $p^2\ll (mc)^2$, this 
gives $u^2=c^2+O(1)$, hence $p^2=(mu)^2=(mc)^2+O(1)$ and
$p_0=\sqrt{(mc)^2+\p^2}=mc+O(c^{-1})$. Therefore,
\[
\frac{p_0^2-(mc)^2}{2m}=(p_0-mc)\frac{p_0+mc}{2m}=(p_0-mc)(c+O(c^{-1}))
=p_0c-mc^2+O(c^{-2}),
\]
\[
\bary{lll}
L&=&\D J\Big(\frac{p^2-(mc)^2}{2m}\Big)
 =J\Big(\frac{p_0^2-(mc)^2}{2m}\Big)-J\Big(\frac{\p^2}{2m}\Big)\\
&~&\\
&\D\approx& J(p_0c-mc^2)-J(\p^2/2m)=E-J(\p^2/2m)
\eary
\]
with the rest frame energy $E=J(p_0c-mc^2)$.
Thus, the constraint Schr\"odinger equation reduces in the 
nonrelativistic limit to the standard Schr\"odinger equation 
for a multiparticle system with Hamiltonian 
\[
H=J(\p^2/2m)+\sum_{a<b}\widehat{U^{ab}}(\x^b-\x^a).
\]
Arbitrary local pair interactions can be obtained
in the nonrelativistic limit by choosing $U^{ab}$ appropriately
(and in a non-unique way). 
For larger kinetic energies, the potential in position space acquires 
additional, nonlocal terms (that can be approximated using derivatives
in the interaction). Thus we have a flexible covariant theory 
with a good nonrelativistic limit.

In particular, from the covariant potential \gzit{e.pairint} 
with pair interactions of the form \gzit{e.relcoul}, we recover 
in the nonrelativistic limit the {\bf standard nonrelativistic
multiparticle dynamics} in the presence of electromagnetic and 
gravitational forces.

\section{Golden opportunities}

   \hfill\parbox[t]{9.5cm}{\footnotesize 

{\em
Behind it all is surely an idea so simple, so beautiful,
that when -- in a decade, a century, or a millennium -- we grasp it,
we will all say to each other, how could it have been otherwise?
}

John Archibald  Wheeler, 1987 \cite{Whe}

\bigskip
{\em
Eine mathematische Theorie ist nicht eher als vollkommen anzusehen, 
als bis du sie so klar gemacht hast, da{\ss} du sie dem ersten Manne 
erkl\"aren k\"onntest, den du auf der Stra{\ss}e triffst.
}

David Hilbert, 1900 \cite{Hil}\\

}\nopagebreak

I do not know whether the perfection requested by Hilbert 
can be achieved in deep theories. But, having discovered the 
unexpected beauty of the present approach, I hope that 
the insights presented will contribute to the perfection of 
quantum field theory.

In 1972, Freeman {\sc Dyson} \cite{Dys} gave a lecture called 
``{\em Missed opportunities}'', where he talked 
``{\em about the contribution that mathematics ought to have made}''
to physics ``{\em but did not}''.
I believe the present contribution widely opens the door for
mathematicians to contribute to quantum field theory, and creates
golden opportunities for those interested in mathematical physics. 

The present setting gives a mathematically consistent point of view 
from which to study the laws of physics, which complements the point 
of view taken by past history.
On the new basis, it is likely that scientists will resolve in 
the near future the most basic challenges current theoretical 
physics poses to mathematicians and mathematical physicists:
\begin{itemize}
\item the existence of QED and derivation of its properties,
\item bound states and resonances in quantum field theories,
\item a unified quantum field theory of all forces of nature,
\item the existence and mass gap in quantum Yang-Mills theory
-- one of seven Clay millenium prize problems \cite{Clay}, 
a golden opportunity in the most concrete sense.
\end{itemize}


\section{Thanks}

It is a great pleasure for me to be able to participate in the 
revelation of the laws the Creator has built into our universe. 
I want to thank {\sc God}
for the call, vision, open-mindedness, patience, persistence and joy 
I got (and needed) for going successfully through the journey in the 
platonic world of precise ideas (that, for a long time, appeared to me 
all too foggy in the regions where quantum field theory is located) 
that lead to the results presented here.

I also want to thank the maintainers of (and the contributors to) the 
Los Alamos National Laboratory {\sc e-Print archive} 
for this wonderful on-line source containing most physics manuscripts 
of the last few years. 
It saved me many hours of work by giving me quick access to the many 
thousands of papers that I glanced at, leaved through, or read more 
thoroughly while searching for the path to success.

I'd like to thank Dr.\ Hermann {\sc Schichl} (Wien) for many 
discussions on various pieces of the puzzle that helped me to clarify 
my thoughts. Thanks also 
to Prof.\ Peter {\sc Michor} (Wien) who pointed me to the book
by {\sc da Silva \& Weinstein} \cite{daSW} on Poisson algebras,
to Prof.\ Walter {\sc Thirring} (Wien) for his treatise on mathematical 
physics which I used over and over again,
to Prof.\ Gerhard {\sc Ecker} (Wien) for useful discussions on quantum 
field theory long ago, and 
to Prof.\ Hartmann {\sc R\"omer} (Freiburg), who introduced me 
many years ago to the idea that an elementary particle `is'
\cite[p.149]{Ste}
an irreducible unitary representation of the Poincar\'e group.

\bigskip


\end{document}